\RequirePackage{fix-cm}

\documentclass[twocolumn]{svjour3}          % twocolumn
\smartqed  % flush right qed marks, e.g. at end of proof
\usepackage{graphicx}

\usepackage{latexsym}
\usepackage{amsmath}
\usepackage{amssymb}
\usepackage{bm}
\usepackage{subcaption}
\newcommand{\sign}{\operatorname{sign}}

\begin{document}

\title{Asymmetric VI-NES with dry friction: An impact map approach 
}

\author{Balkis Youssef \and A. Yassine Karoui \and Remco I. Leine %etc.
}

\institute{Institute for Nonlinear Mechanics, University of Stuttgart, Germany
              \email{balkis.youssef@inm.uni-stuttgart.de}           %  \\
}

\date{Received: date / Accepted: date}

\maketitle

\begin{abstract}
This paper examines the dynamics of a vibro-impact nonlinear energy sink (VI-NES) using a generalized impact map approach. The study incorporates asymmetry and dry friction, reflecting realistic conditions. The proposed method identifies all periodic solutions and determines their stability, and is applicable to various VI-NES configurations, including horizontal and vertical orientations. Numerical results validate prior findings for symmetric frictionless cases and extend them to include frictional and asymmetric dynamics, providing a powerful tool for optimizing the performance of VI-NES in vibration mitigation.
\keywords{VI-NES \and impact map \and friction \and asymmetry \and periodic solutions \and nonlinear normal modes \and targeted energy transfer.}

\end{abstract}

\section{Introduction}
Effective vibration mitigation is an essential requirement for many dynamical structures exposed to harmful oscillations. Without adequate damping, structures with low damping levels can undergo large-amplitude vibrations even under moderate external excitations, potentially resulting in structural fatigue, damage, or failure.
Vibro-impact nonlinear energy sinks (VI-NES) are proven to be effective passive devices for dissipating unwanted vibrational energy in mechanical structures through repeated impacts. Their potential in practical applications has led to extensive research, particularly in aerospace, automotive, and civil engineering, where vibrations can be induced by impulsive loading, seismic excitation, flutter, or collisions \cite{lu2018particle,rahman2015performance,chen2013tuned,lu2018particle2}.
A defining feature of nonlinear energy sinks, including the vibro-impact type, is that their working principle relies on the phenomenon of targeted energy transfer (TET). TET describes the one-way, irreversible transfer of vibrational energy from a primary structure to a local attachment, where the energy is locally dissipated \cite{lu2018particle,gendelman2001transition,vakakis2001inducing}. This energy transfer, also referred to as energy pumping, is related to the occurrence of $p:k$ resonance capture between the NES and the primary structure \cite{lu2018particle,gendelman2001energy,vakakis2001energy}. In other words, TET takes place when the local nonlinear attachment undergoes an internal resonance with one of the modes of the main structure, whether in response to free unforced vibrations induced by transient loading (shock, seismic excitation) \cite{al2013numerical,bapat1985single}, forced resonant vibrations \cite{gendelman2015dynamics,gourc2015targeted} or self-excited vibrations \cite{chatterjee1996impact}.
Due to its proven effectiveness in mitigating resonant vibrations without requiring external control, VI-NES has been widely studied in various configurations (e.g.\ single-sided VI-NES, double-sided VI-NES, VI-NES coupled to cubic nonlinearities and rotatory VI-NES) \cite{wu2023targeted,al2021comparison,farid2021dynamics,saeed2020rotary,al2013numerical,wang2016numerical}. Most studies have focused on symmetric configurations of VI-NES and specific response regimes \cite{vakakis2001inducing,theurich2022predictive,vakakis2008nonlinear,gendelman2015dynamics,gourc2015targeted}, but many aspects of their dynamics remain unexplored. Analytical, numerical, and experimental studies \cite{al2013numerical,qiu2019design,pennisi2017experimental,youssef2021complete} have primarily examined the most relevant response regime—two symmetric impacts per excitation cycle near primary resonance for a frictionless symmetric VI-NES. Obviously, this configuration serves as a logical starting point, as symmetry simplifies the analysis by avoiding unnecessary complexities and isolates the effects of the key nonlinear phenomenon of interest, namely impacts, on the system's behavior. 
However, these simplifying assumptions also introduce limitations, as they do not account for the complexities arising from real-world conditions, such as friction and asymmetry.
Asymmetry, typically induced by gravitational effects in vertical configurations, as well as friction, has been investigated to some extent \cite{li2025effectiveness,theurich2019effects,li2021importance,wang2016numerical}. However, these studies often relied on simplified friction models and a symmetric design with the same coefficients of restitution and friction coefficients, which do not fully reflect real-world scenarios. Moreover, while higher-period response regimes have been identified numerically at elevated excitation levels \cite{li2017optimization,li2021importance}, their dynamics and stability have not been systematically explored. 
This work addresses these issues by employing a generalized impact map approach \cite{leine2012global,liu2023maps} to study the dynamics of an asymmetric VI-NES under dry friction. This method extends previous findings by enabling the computation of all possible periodic solutions, regardless of complexity, and analyzing their stability. The approach is versatile, and can be applied to various VI-NES configurations in horizontal, vertical, or tilted orientations. Incorporating realistic asymmetry conditions enables a reassessment of the optimality criteria for the VI-NES performance \cite{wu2023targeted,li2025effectiveness,qiu2019design,youssef2021complete}.
The paper starts with a description of the model of a VI-NES attached to a linear oscillator (LO) and reduces the problem to a one-degree-of-freedom system focused on the absorber dynamics. Next, the generalized impact map is introduced, detailing how periodic motions with predefined impact sequences and periods are determined, followed by a brief discussion of the stability analysis of these solutions. Finally, the numerical results are presented, validating prior findings for symmetric and frictionless cases, and extending them to a more general framework.
These insights are crucial for practical applications, such as in vibration suppression and energy harvesting systems, where the control of impact dynamics and energy transfer is key. The results also lay the groundwork for future experimental validation, allowing for the development of optimized designs that account for both friction and asymmetry in real-world conditions. Further studies could focus on investigating the effects of varying friction coefficients and exploring the behavior of more complex systems with multiple degrees of freedom.

\section{System description}
\begin{figure}[b]
	\centering
	\begin{subfigure}{0.475\textwidth}
		\includegraphics[width=\linewidth]{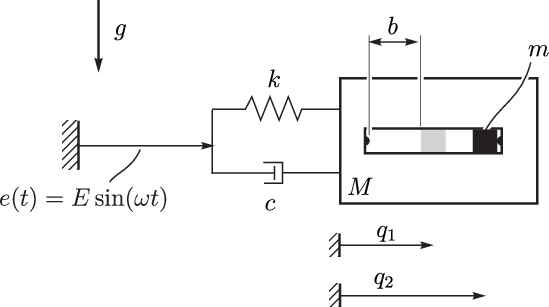}
		\caption{Model of VI-NES attached to a LO.}	\label{fig:model}
	\end{subfigure}
	\begin{subfigure}{0.475\textwidth}	
		\includegraphics[width=\linewidth]{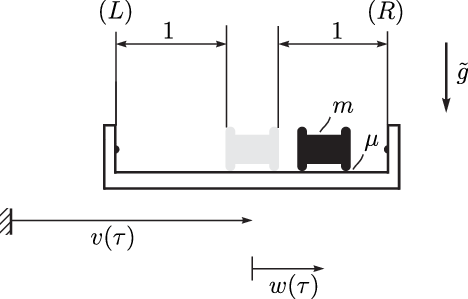}
		\caption{Simplified model.}
		\label{Fig: reduced model}
	\end{subfigure}	
	\caption{Illustration of the harmonically forced mechanical model .}
\end{figure}

The system studied in this paper is an extension of the one previously examined in \cite{youssef2021complete}, which has been frequently used as an illustrative example in multiple studies. The original configuration, depicted in Figure (\ref{fig:model}), consists of a VI-NES coupled to a linear oscillator (mass $M$, stiffness $k$, damping $c$) that is subjected to an external harmonic base excitation, denoted by $e(t)$. The VI-NES comprises a small auxiliary mass $m$, modeled as a particle moving within a cavity of width $2b$ embedded in the primary structure. The coordinates $q_1(t)$ and $q_2(t)$ represent the absolute displacements of the primary mass $M$ and the auxiliary mass $m$, respectively, whereas $t$ represents the dimensioned time in $\left[s\right]$.
During the oscillatory motion of the primary structure, the auxiliary mass $m$ undergoes dry friction and can impact the internal side walls of the cavity. This takes place under the assumption of a closed contact between the auxiliary mass and the lower surface of the cavity at all times. Obviously, for the two masses to be in contact on either side, the condition $|q_1-q_2|=b$ must be fulfilled. In this paper, phases of persistent contact are not considered, i.e. phases where the left or right contact is closed during a non-zero time interval. The impacts occurring on the left (L) and right (R) cavity walls are characterized by distinct Newtonian coefficients of restitution $r_L$ and $r_R$, respectively. Additionally, the frictional force acting on $m$ is characterized by different friction coefficients for rightward and leftward sliding, denoted by $\mu_R$ and $\mu_L$, respectively. In this context, the term \textit{asymmetric} indicates that the energy absorption and dissipation characteristics of the absorber differ depending on the direction of motion. It is important to note that, for the horizontal setup, asymmetry results solely from the friction coefficients and impact characteristics differing between leftward and rightward motion. In a tilted or vertical configuration, the asymmetry is further amplified due to the gravitational effect acting on the absorber.\\
The motion can be divided into two distinct phases: First, the phase where the absorber is moving inside the externally excited cavity, i.e.\ between impacts, described by the following equations:
\begin{align}
	\begin{split}
&\quad\textbf{R $\rightarrow$ L}: \;	|q_1-q_2|<b \, , \; \sign(\dot{q}_1-\dot{q}_2)= +1:\\
		&\quad \left\{
		\begin{array}{rl}
		 M \ddot{q}_1+c \dot{q}_1+k q_1 =& k e(t) + c \dot{e}(t) - \mu_L m g  \\ 
	  m \ddot{q}_2=& \mu_L m g 
		\end{array} \right. 
		\end{split}	\label{eq: R-L q_1 q_2}\\
		\begin{split}
&	\quad	\textbf{L $\rightarrow$ R}: \;	|q_1-q_2|<b \, , \;  \sign(\dot{q}_1-\dot{q}_2)= -1:\\
		&\quad \left\{
		\begin{array}{rl}
			M \ddot{q}_1+c \dot{q}_1+k q_1 =& k e(t) + c \dot{e}(t) + \mu_R m g  \\
			m \ddot{q}_2=& -\mu_R m g  
		\end{array} \right. \label{eq: L-R q_1 q_2}
	\end{split}	
\end{align}
Here, the dot $\left(\dot{-}\right)$ above a coordinate denotes the differentiation with respect to time $t\,\left[s\right]$. The second phase describes a collision between both masses, at impact time instants denoted by $t_n$, at which the $n$-th contact occurs. The impulsive dynamics are obtained with the Newtonian impact law and the principle of momentum conservation:
\begin{align}
	\begin{split}
		&\quad	\textbf{L}: \;	q_1-q_2=b :\\
		& \quad \left\{
		\begin{array}{rl}
	\left(\dot{q}_1^+ -\dot{q}_2^+\right)=&-r_L \left(\dot{q}_1^- -\dot{q}_2^-\right)\\
	 \quad   M \dot{q}_1^+ + m\dot{q}_2^+=&  M \dot{q}_1^- + m\dot{q}_2^-  \, .
		\end{array} \right. 
	\end{split}	 \label{eq:L contact q_1 q_2}\\
	\begin{split}
		&\quad	\textbf{R}: \;	q_1-q_2=-b :\\
	&\quad \left\{
	\begin{array}{rl}
		\left(\dot{q}_1^+ -\dot{q}_2^+\right)=&-r_R \left(\dot{q}_1^- -\dot{q}_2^-\right)\\
		\quad   M \dot{q}_1^+ + m\dot{q}_2^+=&  M \dot{q}_1^- + m\dot{q}_2^-  \, .
	\end{array} \right.
	\end{split}	\label{eq: R contact q_1 q_2}
\end{align}
where $r_L, r_R \in \left[0, \, 1\right]$ and the extreme values $0$ and~$1$ describe an ideal plastic and elastic impact, respectively. The superscripts $(\cdot)^-$ and $(\cdot)^+$ denote the value at the time instant immediately before and after the impact, respectively. To maintain consistency, the same notations for the system parameters and coordinates as in \cite{youssef2021complete} are maintained and the equations are normalized following the same steps to obtain a generalized dimensionless formulation. Consequently, the dimensionless time $\tau$ is introduced as 
\begin{align}
\tau=\omega_0 t  \quad \text{with} \quad \omega_0  = \sqrt{\frac{k}{M}} \, ,
\end{align}
and the new dimensionless coordinates $v$ and $w$ are defined as 
\begin{align}
	v(\tau)&=\frac{1}{b} \left(q_1(\tau)+\epsilon q_2(\tau)\right)\, ,  \\
	w(\tau)&=\frac{1}{b} \left(q_1(\tau)- q_2(\tau)\right) \, ,
\end{align}
where $\epsilon=\frac{m}{M}$ describes the mass ratio between the primary structure and the absorber. Here, the coordinate $v$ represents the motion of the center of mass, while $w$ denotes the normalized relative displacement of the auxiliary mass within the cavity. Considering a harmonic base excitation $e(t)=E \sin\left(\omega t\right)$ acting on the primary structure, the relevant normalized variables and parameters are given by
\begin{equation}
 G=\frac{E}{\epsilon b} \,, \; \Omega=\frac{\omega}{\omega_0}\, , \;
	 \lambda=\frac{c}{m \omega_0} \, , \;	\tilde{g}=\frac{g}{b\omega_0^2}\, .
\end{equation}
Through this normalization, equations (\ref{eq: R-L q_1 q_2})-(\ref{eq: R contact q_1 q_2}) are transformed into 
\begin{align}
	\begin{split}
		&\quad\textbf{R $\rightarrow$ L}: \;	|w|<1 \, , \; w^\prime>0:\\
		&\quad \left\{
		\begin{array}{ll}
v^{\prime \prime}+\epsilon w^{\prime \prime}+\epsilon \lambda v^{\prime}+ v+\epsilon w+\mathcal{O}\left(\epsilon^2\right) = &\\
\qquad \qquad  \epsilon G \sin(\Omega \tau) -\epsilon \mu_L \tilde{g}+\mathcal{O}\left(\epsilon^2\right) \, ,& \\\\
v^{\prime \prime}  -	w^{\prime \prime}=(1+\epsilon) \mu_L \tilde{g}&
		\end{array} \right. 
	\end{split}	\label{eq: R-L v w}\\\nonumber\\
	\begin{split}
	&\quad	\textbf{L}: \;	w=1 :\\
	& \quad \left\{
	\begin{array}{rl}
	&	\left(w^\prime\right)^+	=-r_L \left(w^\prime\right)-\\
& v^+=v^- \, , \quad  	v^{\prime+} =v^{\prime-}  \, .
	\end{array} \right. 
\end{split}	 \label{eq:L contact v w}\\\nonumber\\
	\begin{split}
	&\quad\textbf{L $\rightarrow$ R}: \;	|w|<1 \, , \; w^\prime<0:\\
	&\quad \left\{
\begin{array}{ll}
	v^{\prime \prime}+\epsilon w^{\prime \prime}+\epsilon \lambda v^{\prime}+ v+\epsilon w+\mathcal{O}\left(\epsilon^2\right) =&\\
	\qquad \qquad \epsilon G \sin(\Omega \tau) +\epsilon \mu_R \tilde{g}+\mathcal{O}\left(\epsilon^2\right) \, ,& \\\\
	v^{\prime \prime}  -	w^{\prime \prime}=-(1+\epsilon) \mu_R \tilde{g} &
\end{array} \right. 
\end{split}	\label{eq: R-L v w}\\\nonumber\\
	\begin{split}
	&\quad	\textbf{R}: \;	w=-1 :\\
	& \quad \left\{
	\begin{array}{ll}
	&	\left(w^\prime\right)^+	=-r_R \left(w^\prime\right)-\\
	 & v^+=v^- \, , \quad  	v^{\prime+} =v^{\prime-}  \, .
	\end{array} \right. 
\end{split}	 \label{eq:R contact v w}
\end{align}
where the prime symbol $(\cdot)^\prime$ indicates differentiation with respect to the dimensionless time $\tau$.
This study focuses on cases where the primary structure oscillates (nearly) harmonically with a constant amplitude at a stable steady state. At this state, the system simplifies to an impacting mass moving inside a cavity undergoing a prescribed harmonic excitation $v(\tau)=C \sin\left(\Omega \tau +\gamma\right)$ with a positive constant amplitude $C$ and a constant phase $\gamma$. The primary structure's mass is assumed to be significantly larger than the mass of the absorber, i.e.\ $\epsilon\ll 1$, ensuring that the motion of the primary structure remains unaffected by the impacting mass. As mentioned above, the analysis in this paper is restricted to a horizontal setup, where the contact between the absorber and the cavity surface remains closed at all times. This simplification reduces the problem to a one degree of freedom model, focusing solely on the motion of the absorber within the harmonically excited cavity. However, the methods developed in this work can be easily and straightforwardly extended to tilted or vertical setups, allowing for broader applicability to a wider range of configurations. In such cases, the influence of gravity depends on the tilting angle and modifies only the right-hand side of the equations of motion. These modifications do not affect the proposed methodology. The derivation of the impact map and the subsequent stability analysis remain valid, with only minor adjustements required. Consequently, the relative displacement $w$ remains as the only unknown and its dynamics are governed by the kinematic excitation $v$ acting on the impacting mass $m$. The mechanical model of this simplified system is shown in Figure (\ref{Fig: reduced model}) and the corresponding equation governing the non-impulsive motion between impacts reads
\begin{equation}
	|w|<1 : w^{\prime \prime}= v^{\prime \prime} - (1+\epsilon) \mu_i \tilde{g} \sign(w^\prime) \, , \quad i={\rm R,\, \rm L }
	\label{eq: reduced eq of motion}
\end{equation}
with $i=\rm{R}$ for $w^\prime<0$, i.e.\ when the auxiliary mass is moving from the left to the right (L$\rightarrow$R) and $i=\rm{L}$ for $w^\prime>0$, i.e.\ when it is moving from the right to the left (R$\rightarrow$L). The acceleration of the main structure is given by
\begin{equation}
	v^{\prime \prime}(\tau)=-C\Omega^2 \sin\left(\Omega \tau +\gamma\right) \, .
\end{equation} 
For simplicity, the values of the relative displacement and velocity of the absorber at the time instant $\tau_n$ at which the $n$-th contact occurs, are rewritten with the subscript $(\cdot)_n$. The contact condition and the impact law read
\begin{equation}
	|w\left(\tau_n\right)|=1: \left\{
	\begin{array}{ll}
		w_n=	-1 & {(\rm{R})} ,\,  \left(w_n^\prime\right)^+=-r_R \left(w_n^\prime\right)^- \, \\
		w_n=	+1 & {(\rm{L})},\,  \left(w_n^\prime\right)^+=-r_L \left(w_n^\prime\right)^- \,
	\end{array} \right. \, .
	\label{eq: impact and contact cond}
\end{equation}
A possible way to combine the non-impulsive and impulsive dynamics of the impacting absorber is to express the dynamics using the framework of nonsmooth dynamics to describe contact, impact and friction, ensuring that penetration is prevented, adhesion between the masses after contact is avoided and frictional forces are accounted for. When formulated as measure differential inclusions \cite{leine2007stability}, they allow for the application of the Moreau time-stepping scheme for numerical integration, which forms the basis for all numerical simulations presented in Section \ref{section: numerical results}.
%****************************************************************************
\section{Impact map} \label{Section: Impact map}
Having discussed how to describe the system in continuous time through its equations of motion, this section addresses a way to depict the motion in discrete time with the so-called \textit{impact map}. The term impact map refers to a discrete mapping relating state vectors at successive post-impact time instants with each other. 
An impact event can be characterized by two coordinates: the collision time instant $\tau_n$ and the relative velocity $w^\prime (\tau_n)$ either immediately before or immediately after the impact. For a compact and convenient formulation that ensures that the results remain directly comparable to previous studies, the characterization of the $n$-th impact event is done through a new dimensionless time variable $\Psi_n$ along with the post-impact absolute velocity of the absorber, denoted by $B_n$, defined as follows
\begin{align}
	\begin{split}
\Psi_n=&\Omega \tau_n + \gamma \, ,\\
B_n=&\left(v^\prime_n\right)^+-\left(w^\prime_n\right)^+=C \Omega\cos(\Psi_n) -\left(w^\prime_n\right)^+ 
	\end{split}
 \label{eq:definition Psi B}
\end{align}
Both of these variables are sampled at each impact event, forming a state vector $\mathbf{x}_n=\left(\Psi_n,\, B_n\right)^\mathrm{T}$.
In the case of two alternating impacts per period (2IPP), two possible sequences can occur: either the absorber starts on the right side, undergoes its first impact on the left, followed by a second impact on the right (RL/LR), or the absorber starts on the left side, impacts on the right side first, and then hits the left side (LR/RL). In the asymmetric case, the direction of the mass’s motion within the cavity determines the characteristics and sign of the friction force acting on it. For the impact sequence LR/RL, representing two impacts per cycle, the post-impact state vector $\mathbf{x}_n=\left(\Psi_n,\, B_n\right)^\mathrm{T}$ characterizes the impact, that already occurred within the previous cycle, on the left side. At $\tau_{n+1}$, the absorber reaches the right side, and the impact parameters are updated in the state vector $\mathbf{x}_{n+1}$. After one complete period of excitation, the second impact occurs on the left side, and the corresponding state vector is denoted by $\mathbf{x}_{n+2}$. The impact map for this type of motion, i.e.\ (LR/RL), is a discrete mapping between two state vectors at the same left side, $\mathbf{x}_{n}$ and $\mathbf{x}_{n+2}$, and is defined as
\begin{equation}
	\mathbf{G}\left(\mathbf{x}_n,\, \mathbf{x}_{n+2}\right)=
	\begin{pmatrix}
		\mathbf{G}_{\rm{LR}} \left(\mathbf{x}_n,\, \mathbf{x}_{n+1} \right)\\
		\mathbf{G}_{\rm{RL}} \left(\mathbf{x}_{n+1},\, \mathbf{x}_{n+2} \right)\\
	\end{pmatrix}
	=\mathbf{0} \, , \label{eq: G 2 ipp}
\end{equation}
where $\mathbf{G}_{\rm{LR}}$ and  $\mathbf{G}_{\rm{RL}}$ are the transformations of the post-impact state vectors from the left to the right side, and then from the right side back to the left to close the (LR/RL) sequence. In a more general case of $l$ impacts per $k$ cycles ($1$:$k$ internal resonance with $l$ impacts), the impact sequence becomes more complex compared to the typical two alternating impacts. Specifically, the system can experience impacts on the same side or alternate between the left and right sides within a given cycle. In the following, the fundamental transformations, referred to as elementary mappings and denoted by $\mathbf{G}_{\rm{RL}}$, $\mathbf{G}_{\rm{LR}}$, $\mathbf{G}_{\rm{RR}}$ and $\mathbf{G}_{\rm{LL}}$, are derived in order to obtain the discrete mapping $\mathbf{G}$ for any given impact sequence.
\subsection{Elementary mappings for alternating impacts (LR/RL)}
Considering the case of two alternating impacts per period (2IPP) impact sequence (LR/RL), the expressions for the displacement and velocity of the absorber between the first two consecutive impacts (LR) follow from (\ref{eq: reduced eq of motion}), which reads for $i=\rm{R}$
\begin{equation}
	w^{\prime \prime}= v^{\prime \prime} + (1+\epsilon) \mu_R \tilde{g} \, . \label{eq: rightward motion m}
\end{equation}
Double integration over the open time interval $\left(\tau_n,\, \tau_{n+1}\right)$ and using the corresponding contact and impact equations from (\ref{eq: impact and contact cond}), yields the following implicit equations
\begin{align}
	\begin{split}
	\left(w_{n+1}^\prime\right)^+ =&-r_R\left(w_{n}^\prime\right)^+ -r_R \mu_R \tilde{g}\left(1+\epsilon\right) \left(\tau_{n+1}-\tau_n\right)\\
	& -r_R C \Omega \left(  \cos \left(\Omega \tau_{n+1}+\gamma\right) -\cos\left(\Omega \tau_n +\gamma\right)\right) \, , 
\end{split} \label{eq: velocity equation LR}
\end{align}
\begin{align}
\begin{split}
	w_{n+1}=& w_{n} +C \left(\sin \left(\Omega \tau_{n+1}+\gamma\right)-\sin\left(\Omega \tau_n +\gamma\right)\right) \\ 
	& +\left(	\left(w_{n}^\prime\right)^+-C \Omega \cos\left(\Omega \tau_n +\gamma\right)\right)\left(\tau_{n+1}-\tau_n\right) \\
	&+\frac{\left(1+\epsilon\right)}{2}\mu_R \tilde{g}\left(\tau_{n+1}-\tau_n\right)^2 \, , 
\end{split}\label{eq: Time equation LR}
\end{align}
where $w_{n}=1$ (L) and $w_{n+1}=-1$ (R). Similarly, for the second part of the motion, i.e.\ moving from the right to the left side, the acceleration (\ref{eq: reduced eq of motion}) for $i=\rm{L}$ reads as 
\begin{equation}
	w^{\prime \prime}= v^{\prime \prime} - (1+\epsilon) \mu_L \tilde{g} \, , \label{eq: leftward motion m}
\end{equation}
which after double integration over the open time interval $\left(\tau_{n+1},\, \tau_{n+2}\right)$ yields
\begin{align}
	\begin{split}
	\left(w_{n+2}^\prime\right)^+ =-r_L\left(w_{n+1}^\prime\right)^+ +r_L \mu_L \tilde{g} \left(1+\epsilon\right) \left(\tau_{n+2}-\tau_{n+1}\right) & \\
	-r_L C \Omega \left(  \cos \left(\Omega \tau_{n+2}+\gamma\right)-\cos\left(\Omega \tau_{n+1} +\gamma\right)\right)&  \, ,
\end{split} \label{eq: velocity equation RL}
\end{align}
\begin{align}
\begin{split}
	w_{n+2}=w_{n+1} +C \left(\sin \left(\Omega \tau_{n+2}+\gamma\right)-\sin\left(\Omega \tau_{n+1} +\gamma\right)\right) & \\
	+\left(	\left(w_{n+1}^\prime\right)^+-C \Omega \cos\left(\Omega \tau_{n+1} +\gamma\right)\right)\left(\tau_{n+2}-\tau_{n+1}\right)& \\
	 -\frac{\left(1+\epsilon\right)}{2}\mu_L \tilde{g}\left(\tau_{n+2}-\tau_{n+1}\right)^2 &
	\, ,
\end{split}\label{eq: Time equation RL}
\end{align}
where $w_{n+1}=-1$ (R) and $w_{n+2}=1$ (L).
Using (\ref{eq:definition Psi B}) allows to express the above derived equations in terms of the post-impact coordinates.
Accordingly, the post-impact state vectors $\mathbf{x}_n$ and $\mathbf{x}_{n+1}$, which characterize the $n$-th and $n$+1-th impacts as the absorber moves from the left to the right side, are related through the elementary transformation $\mathbf{G}_{\text{LR}}$, given by Table~(\ref{eq: G LR }), where $\tilde{G}=\frac{ \tilde{g}(1+\epsilon)}{\Omega}$. Similarly, the post-impact coordinates relating the $n$+1-th and $n$+2-th impacts within the same cycle, as the absorber moves from the right to the left side, are related through the elementary transformation $\mathbf{G}_{\text{RL}}$ as given by Table~\ref{eq: G RL }).  
\begin{table*}[ht]
	\centering
	\begin{subtable}[t]{\textwidth}  % Adjust width
		\centering
		\begin{align*}
			&&	\mathbf{G}_{\text{LR}}\left(\mathbf{x}_n,\, \mathbf{x}_{n+1}\right)=
			\begin{pmatrix}
				B_{n+1} +r_R B_n - \left(1+r_R\right) C\Omega \cos\Psi_{n+1}-r_R \mu_R \tilde{G} \left(\Psi_{n+1}-\Psi_n\right) \\
				-\frac{B_n}{\Omega}\left(\Psi_{n+1}-\Psi_n\right)+C \left(\sin\Psi_{n+1}-\sin\Psi_n\right)+\frac{\mu_R\tilde{G}}{2 \Omega}\left(\Psi_{n+1}-\Psi_n\right)^2+2 
			\end{pmatrix}=\mathbf{0}\, .
		\end{align*}
		\caption{}\label{eq: G LR }
	\end{subtable}
	\hfill
	\begin{subtable}[t]{\textwidth}
		\centering
		\begin{align*}
		&&	\mathbf{G}_{\text{RL}}\left(\mathbf{x}_{n+1}, \mathbf{x}_{n+2}\right) =
			\begin{pmatrix}
				B_{n+2} +r_L B_{n+1} - \left(1+r_L\right) C\Omega \cos\Psi_{n+2}+r_L\mu_L \tilde{G} \left(\Psi_{n+2}-\Psi_{n+1}\right) \\
				-\frac{B_{n+1}}{\Omega}\left(\Psi_{n+2}-\Psi_{n+1}\right)+C \left(\sin\Psi_{n+2}-\sin\Psi_{n+1}\right)-\frac{\mu_L \tilde{G}}{2\Omega}\left(\Psi_{n+2}-\Psi_{n+1}\right)^2-2 
			\end{pmatrix} = \mathbf{0} \, .
		\end{align*}
		\caption{}\label{eq: G RL }
	\end{subtable}
	\caption{Elementary mappings for alternating impacts.}
	\label{tab:equations G_LR/G_RL}
\end{table*}

\subsection{Elementary mappings for consecutive impacts on the same side (LL/RR)}
While the previous section introduced the key equations necessary to describe the structure of the elementary transformations, this section extends the analysis to cases where the absorber impacts the same side twice.  These transformations can be obtained analogously by adapting (\ref{eq: velocity equation LR})-(\ref{eq: Time equation RL}). For instance, to establish the relationship between two successive impacts on the left wall ($w_n=w_{n+1}=1$) without the mass reaching the right wall ($w=-1$), the motion of the absorber is constrained between $w=+1$ and a turning point $w=w_t$ where $-1<w_t<1$. Due to the asymmetry in friction coefficients for leftward and rightward motion, the dynamics between successive impacts can be divided into two distinct phases:
\begin{itemize}
\item \textbf{Phase 1:} Rightward motion ($i=\rm{R}$), occurring between $\tau_n$ and $\tau_t<\tau_{n+1}$, governed by (\ref{eq: rightward motion m}).
\item \textbf{Phase 2:} Leftward motion ($i=\rm{L}$), occurring between $\tau_t$ and $\tau_{n+1}$, governed by (\ref{eq: leftward motion m}).
\end{itemize}
The introduction of the turning point introduces two new unknowns, namely, the switching time $\tau_t$ and the relative position $w_t=w(\tau_t)$. 
In the first phase, starting from the left side at $w_n=+1$, the absorber moves rightward until it reaches the turning point at $\tau_t$, where $-1<w_t<1$ and $w^\prime_t=0$. Integrating (\ref{eq: rightward motion m}) twice over $\left(\tau_{n},\, \tau_{t}\right)$ and applying the velocity condition at $\tau_t$ yields the following implicit equations
\begin{align}
	\begin{split}
		0=&w_n^\prime + C\Omega \left(\cos(\Omega\tau_t+\gamma)-\cos(\Omega\tau_{n}+\gamma)\right) \\
		&+\mu_R \tilde{g}(1+\epsilon)\left(\tau_t-\tau_n\right)\,.
	\end{split} \label{eq: velocity equation L-Turning point}
\end{align}  
\begin{align}
	\begin{split}
		w_{t}=& w_{n} +C \left(\sin \left(\Omega \tau_{t}+\gamma\right)-\sin\left(\Omega \tau_n +\gamma\right)\right) \\ 
		& +\left(	\left(w_{n}^\prime\right)^+-C \Omega \cos\left(\Omega \tau_n +\gamma\right)\right)\left(\tau_{t}-\tau_n\right) \\
		&+\frac{\left(1+\epsilon\right)}{2}\mu_R \tilde{g}\left(\tau_{t}-\tau_n\right)^2 \, .
	\end{split}\label{eq: Time equation L-Turning point}
\end{align}
Solving equations (\ref{eq: velocity equation L-Turning point})-(\ref{eq: Time equation L-Turning point}) yields $\tau_t$ and $w_t$, which are then substituted into the equations governing the subsequent phase, i.e.\ for $\tau \in (\tau_t, \tau_{n+1}) $ as the absorber returns to the left wall. Similarly, integrating (\ref{eq: leftward motion m}) twice over the interval $\left(\tau_{t},\, \tau_{n+1}\right)$ results in
\begin{align}
	\begin{split}
		\left(w_{n+1}^\prime\right)^+ =&-r_L C \Omega \left(  \cos \left(\Omega \tau_{n+1}+\gamma\right)-\cos\left(\Omega \tau_{t} +\gamma\right)\right)  \\
		&  +r_L \mu_L \tilde{g} \left(1+\epsilon\right) \left(\tau_{n+1}-\tau_{t}\right) \, ,
	\end{split} \label{eq: velocity equation Turning point-L}
\end{align}
\begin{align}
	\begin{split}
		w_{n+1}=&w_{t} +C \left(\sin \left(\Omega \tau_{n+1}+\gamma\right)-\sin\left(\Omega \tau_{t} +\gamma\right)\right)  \\
		&-C \Omega \cos\left(\Omega \tau_{t} +\gamma\right)\left(\tau_{n+1}-\tau_{t}\right) \\
		& -\frac{\left(1+\epsilon\right)}{2}\mu_L \tilde{g}\left(\tau_{n+1}-\tau_{t}\right)^2 
		\, ,
	\end{split}\label{eq: Time equation Turning point-L}
\end{align} 
where $w_n=w_{n+1}=1$ (L). Using the coordinates introduced in (\ref{eq:definition Psi B}), the six unknowns in the system of equations (\ref{eq: velocity equation L-Turning point})–(\ref{eq: Time equation Turning point-L}) can be reduced to five by applying the condition $B_t=C\Omega \cos\Psi_t$, which follows from $w^\prime_t=0$. Consequently, the impact map for this impact sequence is given by Table~\ref{eq: G LL }). The elementary mapping for successive impacts on the right wall is derived analogously and is given by Table~\ref{eq: G RR }). 
It is worth noting that the structure of the elementary mapping differs depending on the impact sequence. In particular, the dimensionality of the impact map, and consequently the number of unknowns, varies from case to case.
\begin{table*}[ht]
	\centering
	\begin{subtable}[t]{\textwidth}  
		\centering
		\begin{equation*}
				\mathbf{G}_{\text{LL}}\left(\mathbf{x}_n,\, \mathbf{x}_{n+1}\right)=
			\begin{pmatrix}
				C \Omega \cos\Psi_t-B_n+\mu_R\tilde{G}\left(\Psi_t-\Psi_n\right)\\
				B_{n+1} +r_L C \Omega \cos\Psi_t - \left(1+r_L\right) C\Omega \cos\Psi_{n+1}+r_L \mu_L \tilde{G} \left(\Psi_{n+1}-\Psi_t\right) \\
				C\left(\sin\Psi_{n+1}-\sin\Psi_n\right)-\frac{B_n}{\Omega}\left(\Psi_{t}-\Psi_n\right)-\frac{B_t}{\Omega}\left(\Psi_{n+1}-\Psi_t\right)+\frac{\mu_R\tilde{G}}{2 \Omega}\left(\Psi_{t}-\Psi_n\right)^2-\frac{\mu_L\tilde{G}}{2 \Omega}\left(\Psi_{n+1}-\Psi_t\right)^2
			\end{pmatrix}=\mathbf{0} \, .
		\end{equation*}
		\caption{}	\label{eq: G LL } 
	\end{subtable}
	\hfill
	\begin{subtable}[t]{\textwidth}
		\centering
		\begin{equation*}
		\mathbf{G}_{\text{RR}}\left(\mathbf{x}_n,\, \mathbf{x}_{n+1}\right)=
	\begin{pmatrix}
		C \Omega \cos\Psi_t-B_n-\mu_L\tilde{G}\left(\Psi_t-\Psi_n\right)\\
		B_{n+1} +r_R C \Omega \cos\Psi_t - \left(1+r_R\right) C\Omega \cos\Psi_{n+1}-r_R \mu_R \tilde{G} \left(\Psi_{n+1}-\Psi_t\right) \\
		C \left(\sin\Psi_{n+1}-\sin\Psi_n\right)-\frac{B_n}{\Omega}\left(\Psi_{t}-\Psi_n\right)-\frac{B_t}{\Omega}\left(\Psi_{n+1}-\Psi_t\right)-\frac{\mu_L\tilde{G}}{2 \Omega}\left(\Psi_{t}-\Psi_n\right)^2+\frac{\mu_R\tilde{G}}{2 \Omega}\left(\Psi_{n+1}-\Psi_t\right)^2 
	\end{pmatrix} 	=\mathbf{0} \, .
		\end{equation*}
		\caption{}	\label{eq: G RR } 
	\end{subtable}
	\caption{Elementary mappings for consecutive impacts on the same side.}
	\label{tab:equations G_LL/G_RR}
\end{table*}

\subsection{Mapping structure of periodic orbits}
Having defined the elementary mappings, it is now possible to iterate them in order to describe any impact sequence for a general case of $\mathcal{P}_l^k$-orbits, i.e.\ periodic motions with $l$ impacts per $k$-cycles. In this context, the definition $\mathcal{P}_l^k$-orbits, introduced in \cite{leine2012global}, characterizes an orbit of the impact map $\mathbf{G}$ with a period of $\frac{2\pi}{\Omega} k$ and $l$ impacts occurring within one period of oscillation.
In this case, the sought solutions result from solving a set of nonlinear algebraic equations built as 
\begin{equation}
	\mathbf{G}\left(\mathbf{x}_n,\, \mathbf{x}_{n+l}\right)=
	\begin{pmatrix}
		\mathbf{G}_{m_1}\left(\mathbf{x}_n,\, \mathbf{x}_{n+1}\right) \\
		\vdots\\
		\mathbf{G}_{m_l}\left(\mathbf{x}_{n+l-1},\, \mathbf{x}_{n+l}\right) \\
	\end{pmatrix} =\mathbf{0} \, , \label{eq: G general}
\end{equation}
where $ m_1, \cdots , \, m_l \in \left\{ \rm{LR}, \,\rm{RL}, \,\rm{RR}, \,\rm{LL}\right\}$, with the periodicity condition 
\begin{equation}
	\begin{pmatrix}
		\Psi_{n+l}\\
 B_{n+l}
	\end{pmatrix}
	=\begin{pmatrix}
		\Psi_{n}\\
		 B_{n}
	\end{pmatrix}+\begin{pmatrix}
2k\pi\\
0
\end{pmatrix} \, , \quad k \in \mathbb{N} \, .\label{eq: periodicity condition general}
\end{equation}
The periodicity condition (\ref{eq: periodicity condition general}) follows from the fact that the last impact of the sequence occurs exactly one period after the first impact, i.e.\ $\tau_{n+l}-\tau_n=\frac{2\pi}{\Omega} k$, and since periodicity is characterized by the repetition of the same state after a response period, the post-impact velocities $B_n$ and $B_{n+l}$ must be equal.\\
Depending on the impact sequence of interest and whether it includes impacts on the same side or only alternating impacts, the dimensionality of the resulting system of equations (\ref{eq: G general}) varies. However, in both cases, the system is initially underdetermined, with two more unknowns than equations. Imposing the periodicity condition (\ref{eq: periodicity condition general}) establishes a relationship between the state vector at the beginning of the sequence, $\mathbf{x}_n$, and the state vector after a full period, $\mathbf{x}_{n+l}$. This reduces the number of unknowns by two, ensuring that the system of equations is solvable for any impact sequence, under the assumption that for the chosen parameters a periodic motion with said impact sequence exists.
\subsection{Symmetry condition}
From the definition of the general mapping structure emanates the symmetry condition
\begin{align}
	\tau_{n+\frac{l}{2}}=\tau_n+\frac{\pi}{\Omega} k \, .
\end{align}
Hence, only motions with an even-valued number of impacts per period $l$ can be symmetric.
Symmetric motion can only be observed for a symmetric absorber, i.e.\ for $\mu_R=\mu_L$ and $r_R=r_L$. This can, for instance, be shown for the simplest case of 2 impacts per $k$-cycles motion. The assumption of a sequence with two symmetric impacts implies
that the non-impulsive phases after each impact, i.e. the time intervals in which the impacting mass travels from one side to the other, must be equal. Hence, the second impact of the
sequence occurs precisely after a half response period:
\begin{align}
	\tau_{n+1}=\tau_n+\frac{k\pi}{\Omega} \, .
\end{align}
In order for this symmetry requirement to be fulfilled, the post-impact velocities after both impacts must have the same absolute value and opposed signs, since the impacts occur on opposed sides of the cavity, which leads to the symmetry condition formulated as:
\begin{equation}
	\begin{pmatrix}
		\Psi_{n+1}\\
		B_{n+1}
	\end{pmatrix}
	=\begin{pmatrix}
		\Psi_{n}\\
		-B_{n}
	\end{pmatrix}+\begin{pmatrix}
		k\pi\\
		0
	\end{pmatrix} \, , \quad k \in \mathbb{N} \, .\label{eq: symmetry condition 2IPP}
\end{equation}
Hence, the sought periodic solutions, i.e.$\mathcal{P}_2^k$-orbits, can be calculated by
imposing the periodicity condition (\ref{eq: periodicity condition general}), for $l=2$, and the symmetry condition (\ref{eq: symmetry condition 2IPP}) on the discrete mapping $\mathbf{G}$, given by
\begin{equation}
	\mathbf{G}\left(\mathbf{x}_n,\, \mathbf{x}_{n+2}\right)=
	\begin{pmatrix}
		\mathbf{G}_{LR}\left(\mathbf{x}_n,\, \mathbf{x}_{n+1}\right) \\
		\mathbf{G}_{RL}\left(\mathbf{x}_{n+1},\, \mathbf{x}_{n+2}\right) 
	\end{pmatrix} =\mathbf{0} \, , \label{eq: G 2IPkP}
\end{equation}
and solving for $\mathbf{G}=\mathbf{0}$. 
This narrows down the computation of the periodic solutions to only symmetric $\mathcal{P}_2^k$ -orbits. Depending on the parity of $k$, two cases emanate:
 \begin{itemize}
 \item \textbf{Even-valued }$k$, for which holds
 	\begin{align}
 \left.
\begin{array}{ll}
	\cos (\Psi_{n+1})=&\cos(\Psi_n) \\
		\sin (\Psi_{n+1})=&\sin(\Psi_n)\\
		\cos (\Psi_{n+2})=&\cos(\Psi_n) \\
		\sin (\Psi_{n+2})=&\sin(\Psi_n)
\end{array}
\right \} \; \forall \, k=2p \, , \; p \in \mathbb{N}_0 \label{eq: trigonometric relations k even valued}
 	\end{align}
Substituting the trigonometric relations of (\ref{eq: trigonometric relations k even valued}) in (\ref{eq: G LR }) and (\ref{eq: G RL }) yields the following system of equations:
\begin{align}
B_n \frac{k \pi}{\Omega}=&\mu_R \tilde{G}\frac{(k \pi)^2}{2\Omega}+2  \, ,\label{eq:a}\\
B_n \frac{k \pi}{\Omega}=&\mu_L \tilde{G}\frac{(k \pi)^2}{2\Omega}+2 \, ,\label{eq:b}\\
(1+r_R)C\Omega \cos(\Psi_n)=&-(1-r_R)B_n-r_R\mu_R\tilde{G}k\pi\, ,\label{eq:c}\\
(1+r_L)C\Omega \cos(\Psi_n)=&(1-r_L)B_n+r_L\mu_L\tilde{G}k\pi\, .\label{eq:d}
\end{align}
The system formed by (\ref{eq:a})-(\ref{eq:d}) is unsolvable, also under symmetric conditions. Consequently, for even-valued $k$, a periodic motion with two alternating equispaced impacts per $k$-cycles is physically non-viable.\\
\item \textbf{Odd-valued} $k$, for which holds
 	 	\begin{align}
 		\left.
 	\begin{array}{ll}
 		\cos (\Psi_{n+1})=&-\cos(\Psi_n) \\
 		\sin (\Psi_{n+1})=&-\sin(\Psi_n)\\
 		\cos (\Psi_{n+2})=&\cos(\Psi_n) \\
 		\sin (\Psi_{n+2})=&\sin(\Psi_n)
 	\end{array}
 		\right \} \; \forall \, k=2p+1 \, , \; p \in \mathbb{N}_0\label{eq: trigonometric relations k odd valued}
 	\end{align}
 Again, substituting  (\ref{eq: trigonometric relations k odd valued}) in (\ref{eq: G LR }) and (\ref{eq: G RL }) gives the following system of equations:
 \begin{align}
 	2C \sin(\Psi_n)=-B_n \frac{k \pi}{\Omega}+\mu_R \tilde{G}\frac{(k \pi)^2}{2\Omega}+2 \, , \label{eq:a2}\\
 	2C \sin(\Psi_n)=-B_n \frac{k \pi}{\Omega}+\mu_L \tilde{G}\frac{(k \pi)^2}{2\Omega}+2 \, , \label{eq:b2}\\
 	(1+r_R)C\Omega \cos(\Psi_n)=(1-r_R)B_n+r_R\mu_R\tilde{G}k\pi \, ,\label{eq:c2}\\
 	(1+r_L)C\Omega \cos(\Psi_n)=(1-r_L)B_n+r_L\mu_L\tilde{G}k\pi\, .\label{eq:d2}
 \end{align}
 Clearly, equations (\ref{eq:a2}) and (\ref{eq:b2}) are consistent only if the friction coefficients are identical, i.e.\ $\mu_R=\mu_L=\mu$. In this case, asymmetry can still be preserved by choosing, $r_L\neq r_R$, leading to a unique solution for $B_n$ that exists for a specific excitation amplitude $C$:
 \begin{align} 
	B_n=\mu \tilde{G}\frac{k\pi}{2}  \quad \text{for} \quad C^2=1+\left(\mu\tilde{G}\frac{k\pi}{2\Omega}\right)^2 \, .
\end{align}
If a symmetric configuration is chosen, i.e.\ $\mu_R=\mu_L=\mu$ and  $r_R=r_L=r$, the system of equations simplifes to two equations with two unknown variables $\Psi_n$ and $B_n$. These can be solved by expressing the sine and cosine terms of $\Psi_n$ as 
\begin{align}
	\sin(\Psi_n)&=\frac{1}{C} \left(1-\frac{k\pi}{2\Omega}\left(B_n-\mu \tilde{G}\frac{k \pi}{2}\right)\right) \, ,\\
	\cos(\Psi_n)&=\frac{1}{C}\left(\frac{R}{\Omega}\left(B_n-\mu \tilde{G}\frac{k\pi}{2}\right)+\mu \tilde{G}\frac{k\pi}{2\Omega} \right) \, ,
\end{align}
with 
\begin{align}
	R=\frac{1-r}{1+r}  \, . 
\end{align}
Using the trigonometric identity
\begin{align*}
\sin(\Psi_n)^2+\cos(\Psi_n)^2=1 \, ,
\end{align*}
 yields the polynomial expression 
\begin{align}
	\begin{split}
C^2=&\left(1-\frac{k\pi}{2\Omega}\left(B_n-\mu \tilde{G}\frac{k \pi}{2}\right)\right)^2 \\
&+\left(\frac{R}{\Omega}\left(B_n-\mu \tilde{G}\frac{k\pi}{2}\right)+\mu \tilde{G}\frac{k\pi}{2\Omega} \right)^2 \, ,
	\end{split}  \label{eq: SIM general 2IPKP SYMMETRIC}
\end{align}
that describes the set of all periodic solutions with two symmetric impacts per $k$-periods. 
 \end{itemize}
Therefore, periodic motion with two alternating equispaced impacts per $k$-cycles is physically non-viable for asymmetric configurations.
In the case of symmetry, symmetric $\mathcal{P}_2^k$-orbits only exist for odd-valued $k$. For this scenario, the external excitation amplitude $C$  is expressed by a polynomial in terms of the post-impact velocity $B_n$ according to (\ref{eq: SIM general 2IPKP SYMMETRIC}). The obtained expression does not only depend on the coefficient of restitution $r$ and the friction coefficient $\mu$, but also includes $k$, the dimensionless frequency $\Omega$  and the cavity length $b$ (through $\tilde{G}$) as parameters. It is noteworthy that inserting $\mu=0$, $k = 1$ and $\Omega= 1$ leads to the same expression for the SIM obtained in many previous studies of symmetric frictionless cases (e.g.\ \cite{pennisi2017experimental,youssef2021complete}). 

%****************************************************************************
\section{Linear stability and bifurcation analysis}
Once a given $\mathcal{P}_l^k$-orbit is computed by means of (\ref{eq: G general})-(\ref{eq: periodicity condition general}), its stability and bifurcation behavior can
be determined by analyzing the linearized impact map as in \cite{leine2012global}. This involves introducing small perturbations around the periodic solutions and studying their corresponding propagation through an eigenvalue analysis.
Let $\mathbf{x}^*_{n+i}=\left(\Psi^*_{n+i},\,, B^*_{n+i}\right)^\text{T}$, $i=0 \cdots l$, describe a periodic motion such that the periodicity condition (\ref{eq: periodicity condition general})
\begin{equation}
		\mathbf{x}^*_{n+l}=\mathbf{x}^*_{n}+\begin{pmatrix}
		2k\pi\\
		0
	\end{pmatrix}
\end{equation}
is fulfilled. In a generalized sense, $\mathbf{x}^*_{n}$ is a fixed point as 
\begin{equation}
	\mathbf{G}\left(\mathbf{x}^*_{n},\, \mathbf{x}^*_{n+l}\right)=	\mathbf{G}\left(\mathbf{x}^*_{n},\, \mathbf{x}^*_{n}\right)=\mathbf{0}\, .
\end{equation}
The perturbed state is introduced as 
\begin{equation}
	\begin{split}
	&\mathbf{x}_{n+i}=\mathbf{x}^*_{n+i}+\Delta \mathbf{x}_{n+i} \, , \\
	&\text{with} \quad  \| \Delta \mathbf{x}_{n+i}\| \ll 1 , \quad i=0,\cdots,\,l \, .
	\end{split}
\end{equation} 
The perturbed solutions $\mathbf{x}_{n+i}$ are by definition also solutions of the system and therefore fulfill (\ref{eq: G general}), whereas the fixed points of $\mathbf{G}$ must fulfill (\ref{eq: periodicity condition general}) as well. The expansion of (\ref{eq: G general}) up to the first order is given by Table~\ref{eq: expansion}, where the superscript $\left(\cdot_{m_i}\right)^*$ means that the corresponding transformation, i.e.\ $\mathbf{G}_{m_i}$, is evaluated at the fixed points $\left(\mathbf{x}^*_{n+i-1},\, \mathbf{x}^*_{n+i}\right)$.

\begin{table*}[ht]
	\centering
	\begin{align*}
&&	\mathbf{0} =	\mathbf{G}\left(\mathbf{x}_n,\, 	\mathbf{x}_{n+l}\right)=   
			\begin{pmatrix}
				\mathbf{G}^*_{m_1} +\left(\frac{\partial\mathbf{G}_{m_1}}{\partial \mathbf{x}_{n}}\right)^*	\Delta \mathbf{x}_{n} +\left(\frac{\partial\mathbf{G}_{m_1}}{\partial \mathbf{x}_{n+1}}\right)^*	\Delta \mathbf{x}_{n+1}+ \mathcal{O}\left( \| \Delta \mathbf{x}_{n}\|^2\right)   \\
				\vdots\\
				\mathbf{G}^*_{m_l} +\left(\frac{\partial\mathbf{G}_{m_l}}{\partial \mathbf{x}_{n+l-1}}\right)^* \Delta \mathbf{x}_{n+l-1} +\left(\frac{\partial\mathbf{G}_{m_l}}{\partial \mathbf{x}_{n+l}}\right)^*	\Delta \mathbf{x}_{n+l} +\mathcal{O}\left(\| \Delta \mathbf{x}_{n}\|^2\right)
			\end{pmatrix} \, .
		\end{align*}
	\caption{First-order expansion of the impact map $\mathbf{G}$.}
	\label{eq: expansion}
\end{table*}

Hence, the propagation of the perturbations from the first to the $l$-th impact can be described through $l$ separate mappings, according to
\begin{equation}
	\Delta \mathbf{x}_{n+i} =\mathbf{A}_{i}\,	\Delta \mathbf{x}_{n+i-1}  \label{eq: perturbation i_th impact} \, , \quad i=1 \cdots l \, , 
\end{equation}
where
\begin{equation}
	\mathbf{A}_{i} = - \left(\frac{\partial \mathbf{G}_{m_i}}{\partial \mathbf{x}_{n+i}}\right)^{*^{-1}} \left(\frac{\partial\mathbf{G}_{m_i}}{\partial \mathbf{x}_{n+i-1}}\right)^* \, ,
\end{equation}
which combined yield a linear mapping from the initial perturbation $\Delta \mathbf{x}_{n}$ to the perturbation after one period and $l$ impacts $	\Delta \mathbf{x}_{n+l}$ that reads
\begin{align}
	\begin{split}
	\Delta \mathbf{x}_{n+l}=\mathbf{A}\Delta \mathbf{x}_{n} = &\prod_{i=1}^l  \mathbf{A}_{l+1-i} \Delta \mathbf{x}_{n}\\
=& \mathbf{A}_l \mathbf{A}_{l-1} \cdots \mathbf{A}_{1} \Delta \mathbf{x}_{n}\, .
	\end{split}
 \label{eq: linearized system G}
\end{align}
Hence, linear asymptotic stability of the linearized system (\ref{eq: linearized system G}), describing small perturbations around the periodic solution, is guaranteed if all eigenvalues $\lambda_j$ of $\mathbf{A}$  lie within the unit circle ($|\lambda_{j}|<1\, , j=1,2$). In this case, the fixed point of the nonlinear impact map $\mathbf{G}$ is also asymptotically stable. However, changes in system parameters, such as increasing the external excitation amplitude $C$, can destabilize fixed points, leading to bifurcations. If a real eigenvalue crosses $+1$, either a turning point bifurcation of the fixed point ($=$ fold bifurcation of the periodic solution) or pitchfork bifurcation of the fixed point ($=$ symmetry-breaking bifurcation of the periodic solution) may occur. Notably, symmetry-breaking bifurcations cannot arise in motions with an odd number of impacts, as such motions are inherently asymmetric. In contrast, when a real eigenvalue reaches $-1$, a flip bifurcation of the fixed point ($=$ period doubling bifurcation of the periodic solution) occurs, where the new steady-state motion has twice the period of the original one.
%*******************************************
\section{Numerical results} \label{section: numerical results}
In this section, the numerical results of path continuation and linear stability analysis of $\mathcal{P}_l^k$-orbits are examined by means of bifurcation diagrams. These diagrams illustrate how the absolute value of the post-impact velocity $|B|$ varies with the bifurcation parameter chosen as the squared amplitude $C^2$.
The numerical validation of the proposed approach is conducted in three stages. First, the simplest case of a symmetric horizontal configuration without friction is considered, i.e.\ $\mu_L=\mu_R=0$ and $r_l=r_R$. Next, symmetric friction is introduced with $\mu_L=\mu_R\neq0$. Finally, the numerical results are extended to an asymmetric configuration with friction, where $\mu_L\neq\mu_R$ and $r_L\neq r_R$.
%****** symmetric frictionless Case************
\subsection{Symmetric frictionless case}
\begin{figure}[h]
	\centering
	\begin{minipage}{0.475\textwidth}
		\includegraphics[width=\linewidth]{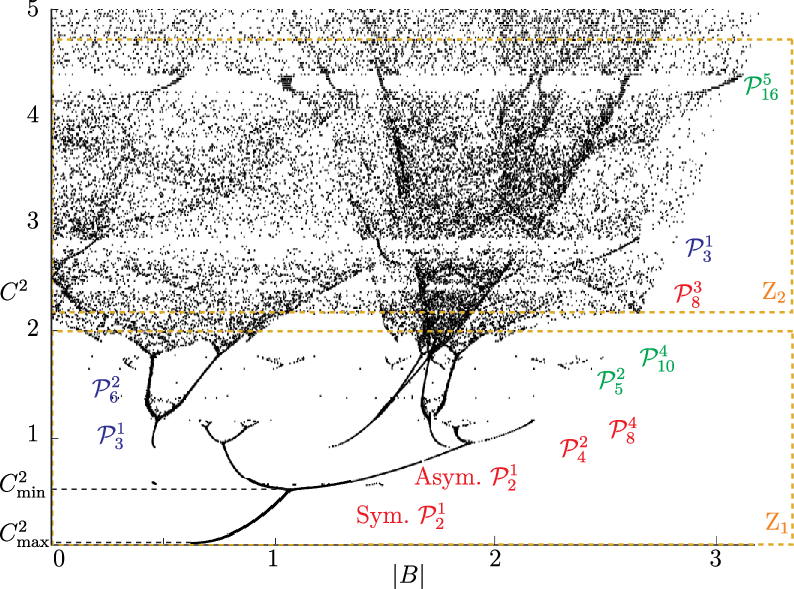}
		\caption{Brute force diagram for $C^2 \in \left[0,\, 5\right]$, $r=0.65$, $\mu=0$, $\Omega=1$ and $k = 1$ .}
		\label{Fig: Brute_force_1}
	\end{minipage}\quad
	\begin{minipage}{0.475\textwidth}
		\includegraphics[width=\linewidth]{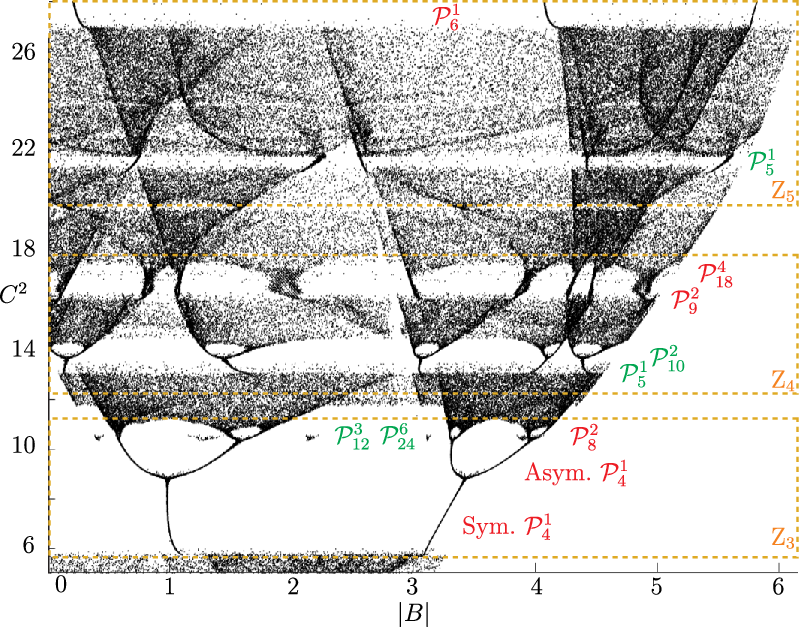}
			\caption{Brute force diagram for $C^2 \in \left[5,\, 27\right]$, $r=0.65$, $\mu=0$, $\Omega=1$ and $k = 1$ .}
		\label{Fig: Brute_force_2}	
	\end{minipage}
\end{figure}

\begin{figure}[h!]
	\centering
	\begin{minipage}{0.475\textwidth}
		\includegraphics[width=\linewidth]{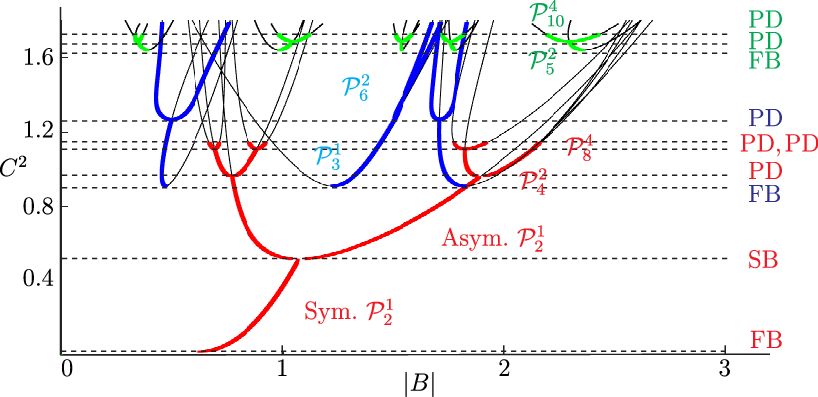}
		\caption{Bifurcation diagram for $C^2 \in \left[0,\, 1.8\right]$, $\mu=0$,\linebreak $r=0.65$, $\Omega=1$ and $k = 1$ obtained via path continuation .}
		\label{Fig: Z1}
	\end{minipage}
\end{figure}

\begin{figure}[h!]
	\centering
	\begin{minipage}{0.475\textwidth}
		\includegraphics[width=\linewidth]{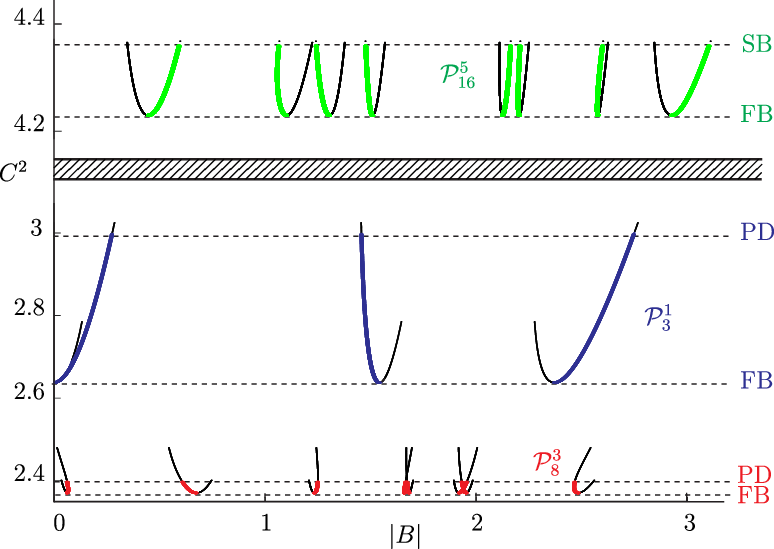}
		\caption{Bifurcation diagram for $C^2 \in \left[2.3,\, 4.4\right]$, $\mu=0$, \linebreak
			 $r=0.65$, $\Omega=1$ and $k = 1$ obtained via path continuation.}
		\label{Fig: Z2}	
	\end{minipage}
\end{figure}

\begin{figure}[h!]
	\centering
	\begin{minipage}{0.475\textwidth}
		\includegraphics[width=\linewidth]{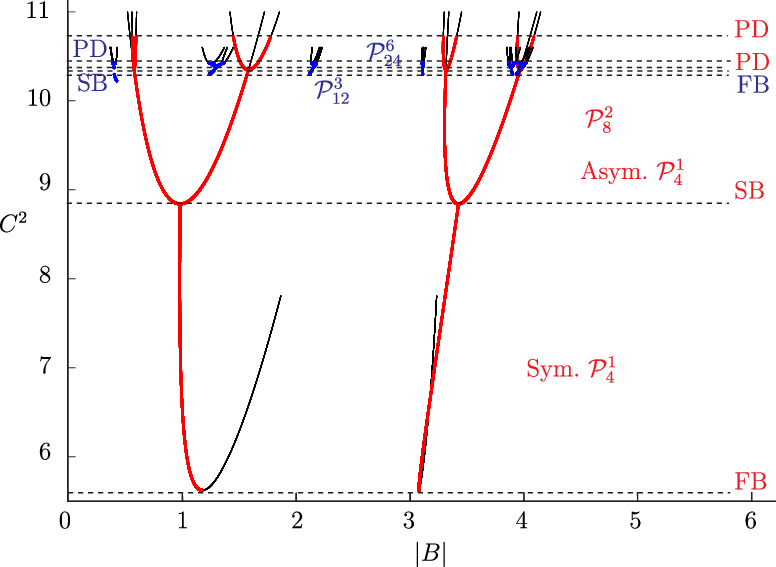}
		\caption{Bifurcation diagram for $C^2 \in \left[5.5,\, 11\right]$, $\mu=0$, \linebreak $r=0.65$, $\Omega=1$ and $k = 1$ obtained via path continuation.}
		\label{Fig: Z3}
	\end{minipage}
\end{figure}

\begin{figure}[h!]
	\centering
	\begin{minipage}{0.475\textwidth}
		\includegraphics[width=\linewidth]{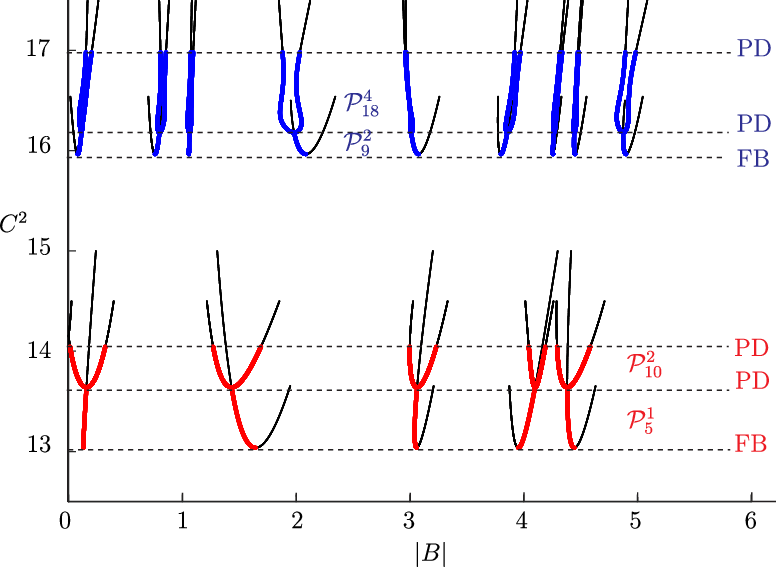}
		\caption{Bifurcation diagram for $C^2 \in \left[12.5,\, 17.5\right]$,\\ 
		$\mu=0$, $r=0.65$, $\Omega=1$ and $k = 1$ obtained via path continuation..}
		\label{Fig: Z4}	
	\end{minipage}
\end{figure}

\begin{figure}[h!]
	\centering
	\begin{minipage}{0.475\textwidth}
		\includegraphics[width=\linewidth]{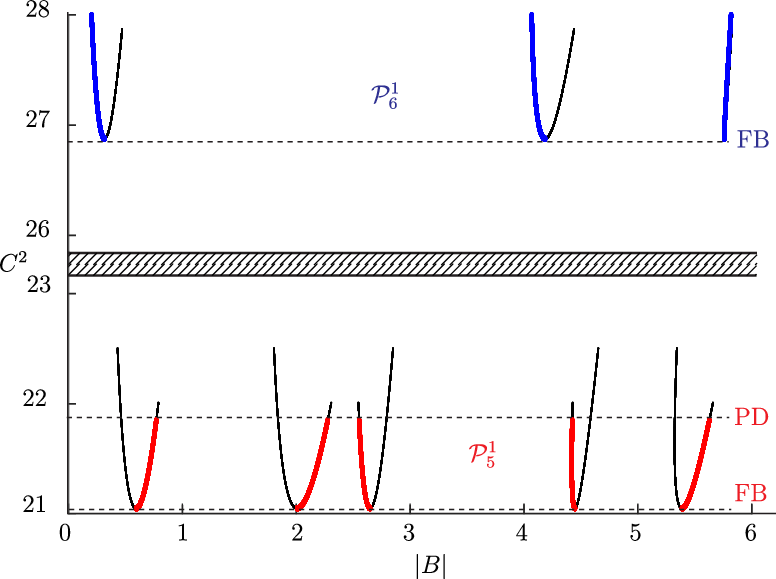}
		\caption{Bifurcation diagram for $C^2 \in \left[21,\, 28\right]$, $\mu=0$,\linebreak $r=0.65$, $\Omega=1$ and $k = 1$ obtained via path continuation.}
		\label{Fig: Z5}	
	\end{minipage}
\end{figure}

This section presents a brute force diagram (Figure (\ref{Fig: Brute_force_1}) and (\ref{Fig: Brute_force_2})) showing the bifurcation scenario with varying $C^2$ for a symmetric setup ($r_L=r_R=r=0.65$, $\mu_L=\mu_R=\mu=0$). These diagrams serve as an independent validation of the impact map approach, confirming the results obtained through continuation methods. A brute force bifurcation diagram is generated by performing multiple time-stepping simulations for different initial conditions at a fixed value of $C^2$, incrementing $C^2$ by $\Delta C^2$ after each simulation, and repeating the process until the entire parameter range is covered. The post-impact velocities recorded after sufficiently long simulations, ensuring the vanishing of transient effects, are plotted as discrete points in the ($|B|, C^2$)-plane.
For periodic motion with two symmetric impacts per $k$-cycle, all corresponding post-impact velocities coincide at a single point in the ($|B|, C^2$)-plane. If the periodic solution has $l$ (even) symmetric impacts, the data points cluster at exactly $\frac{l}{2}$ distinct points. In contrast, for asymmetric periodic motions with $l$ impacts, $l$ distinct points appear for a fixed value of $C^2$. Non-periodic steady-state motion (e.g.\ chaos) leads to distributed points at each $C^2$ level. It is important to note that the brute force bifurcation diagram reveals only the stable branches of periodic solutions, since it is based on time-stepping simulations.
The presented brute force bifurcation diagram is divided in two diagrams for the sake of clarity, as some periodic windows are relatively small and otherwise difficult to distinguish.
Figure (\ref{Fig: Brute_force_1}) first depicts the stable branch of the SIM, followed by stable asymmetric branches with two impacts per cycle. The first two bifurcation points, corresponding to the fold and symmetry-breaking bifurcations, align with the analytically derived values for $r=0.65$, $\mu=0$, $\Omega=1$ and $k = 1$ and are labeled $C^2_{\text{min}}$ and $C^2_{\text{max}}$ in Figure (\ref{Fig: Brute_force_1}). Their corresponding expressions are given by
\begin{equation}
C^2_{\text{min}}=\frac{4 R^2}{(k\pi)^2+4R^2} \; , \quad C^2_{\text{max}}=\frac{16+(2Rk\pi)^2}{((k\pi)^2-4)^2} \, .
\end{equation}
Furthermore, comparing the derived expression for the SIM using the multiple scales method (e.g.\ \cite{gendelman2015dynamics,gourc2015targeted}) with that obtained via the impact map approach highlights the advantages of the presented method. Unlike the MSM, where the SIM depends solely on the coefficient of restitution $r$, the impact map approach incorporates the normalized frequency $\Omega$ and order $k$, extending the results to a broader range of cases.
The symmetric branch of solutions splits into two asymmetric branches, which subsequently undergo a period-doubling bifurcation cascade with a decreasing stability range.
A closer look, shows that before the first period-doubling bifurcation of the $\mathcal{P}_2^1$-orbits, the $\mathcal{P}_3^1$-orbits also exist, which
bifurcate into $\mathcal{P}_6^2$-orbits. These remain stable until another period-doubling
bifurcation occurs, the resulting branches are difficult to distinguish due to their narrow stability range and the coexistence of other motion types within the same $C^2$ range. The brute force results are in congruence with the solution branches obtained via path continuation, which coincide with the periodic windows. For clarity, zoomed views of the path continuation results in regions $Z1-Z5$ are presented in Figures (\ref{Fig: Z1})–(\ref{Fig: Z5}), where stable and unstable branches are marked with colored and thin black lines, respectively.
%***********symmetric Case with friction*************
\subsection{Symmetric case with friction}
\begin{figure}[h]
	\centering
	\begin{minipage}{0.475\textwidth}
		\includegraphics[width=\linewidth]{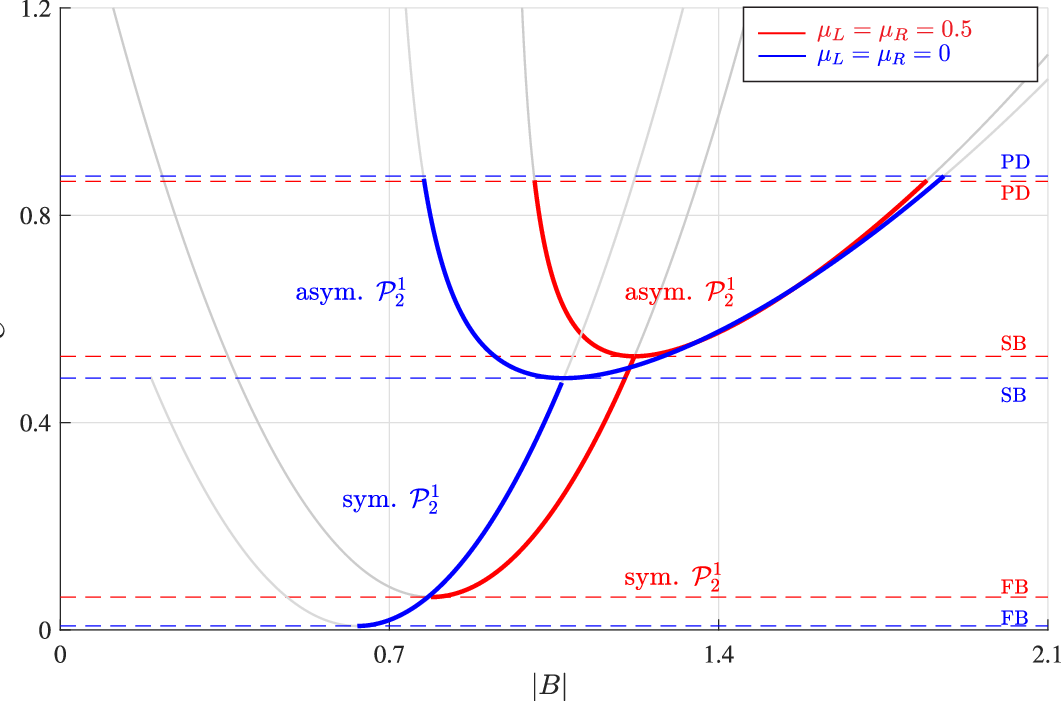}
		\caption{Stability region of the $\mathcal{P}_2^1$-orbits with and without friction for $r_L=r_R=0.76$.}
		\label{Fig: stability region of 2ipp}
	\end{minipage}\quad
	\begin{minipage}{0.475\textwidth}\hspace{-.5cm}
		\includegraphics[width=1.1\linewidth]{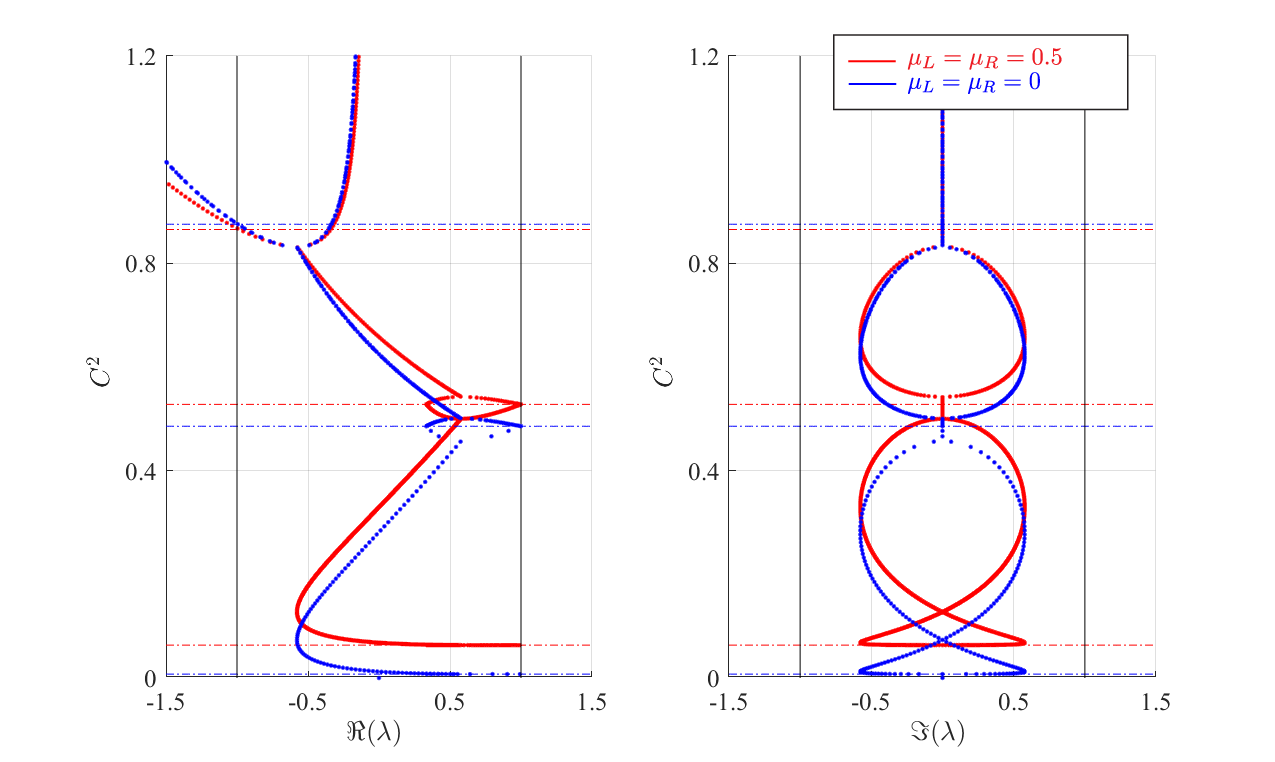}
		\caption{Eigenvalues of the matrix $\mathbf{A}$ for the system with and without friction for $r_L=r_R=0.76$ \linebreak and $\tilde{g}=0.211$. The dashed lines correspond to the excitation levels of the bifurcation points.}
		\label{Fig: eigenvalues 2ipp}	
	\end{minipage}
\end{figure}
This section begins with a bifurcation diagram for the $\mathcal{P}_2^1$-orbits in a symmetric setup with $r_L = r_R = 0.76$, comparing two cases: one with friction ($\mu_L = \mu_R = 0.5$) and one without friction ($\mu_L = \mu_R = 0$). As illustrated in Figures (\ref{Fig: stability region of 2ipp})-(\ref{Fig: eigenvalues 2ipp}), the stable branch of the symmetric $\mathcal{P}_2^1$-orbits for $\mu_R = \mu_L = 0.5$ exhibits topological features similar to those of the slow invariant manifold in the frictionless case.
The symmetric $\mathcal{P}_2^1$-orbits emerge from a fold bifurcation at a higher amplitude compared to the frictionless case, and transition into stable asymmetric branches with two impacts per cycle. As mentioned in Section \ref{Section: Impact map}, this solution branch can be derived analytically, and the stability boundaries are given by
\begin{align}
	C^2_{\min}&= \frac{\left(R+(\frac{\pi}{2})^2\mu\tilde{g}\right)^2}{\frac{\pi^2}{4}+R^2} \,,\\ 
	C^2_{\max}&=\frac{\left(R\pi^2\mu\tilde{g}-4\right)^2}{\left(\pi^2-4\right)^2}+\left(\frac{2R(\pi-R\pi\mu\tilde{g})}{\pi^2-4}+\frac{\pi}{2}\mu\tilde{g}\right)^2 \, .\label{eq: Cbounds}
\end{align}
Hence, friction does not alter the qualitative behavior of the absorber but increases the energy required to activate the VI-NES and initiate targeted energy transfer. Additionally, it narrows the stability range of the $\mathcal{P}_2^1$-orbits, further restricting the operational range of the VI-NES.
As $C^2$ increases, the qualitative behavior follows a similar pattern to the frictionless case, with stable symmetric and asymmetric orbits transitioning through the same bifurcations, ultimately leading to more complex periodic orbits.

%*********Asymmetric case with friction****************
\subsection{Asymmetric case with friction}
\begin{figure}[h]
		\begin{minipage}{0.475\textwidth}
	\centering
	\includegraphics[width=\linewidth]{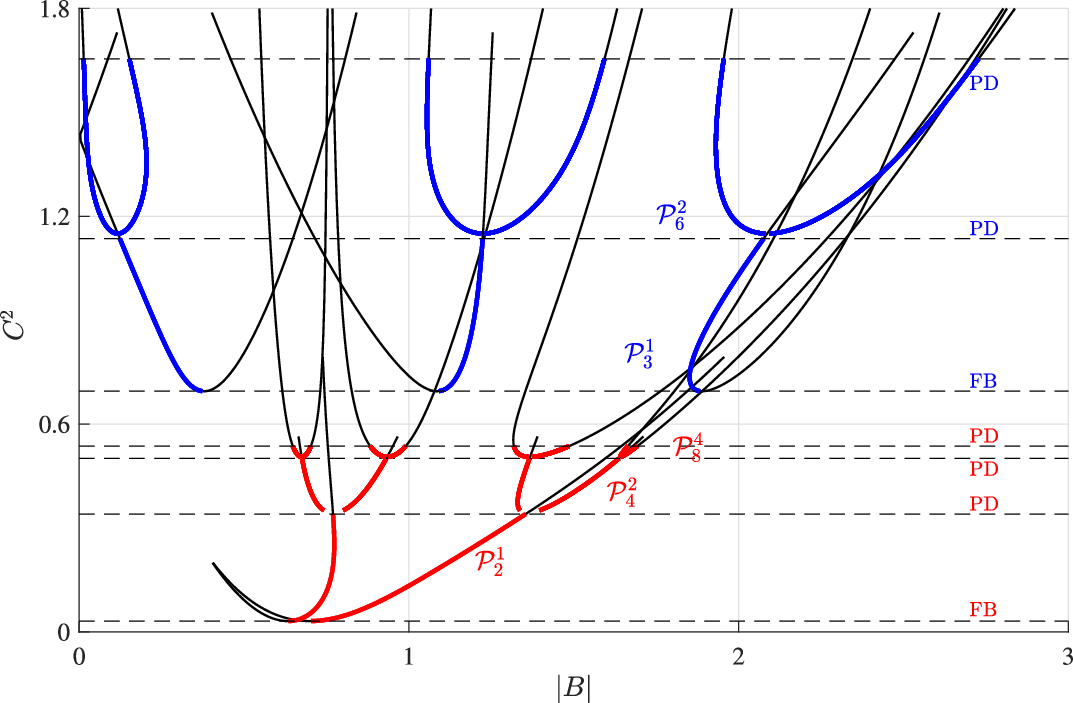}
	\caption{Bifurcation diagram for $r_L=0.56$, $r_R=0.76$, $\mu_R=0.15$, $\mu_L=0.1$ and $\tilde{g}=0.211$.}
	\label{Fig: Bif_diagramm_asymm_final}	
\end{minipage}
\end{figure}

\begin{figure}[h]
	\centering
	\begin{minipage}{0.475\textwidth}
		\includegraphics[width=\linewidth]{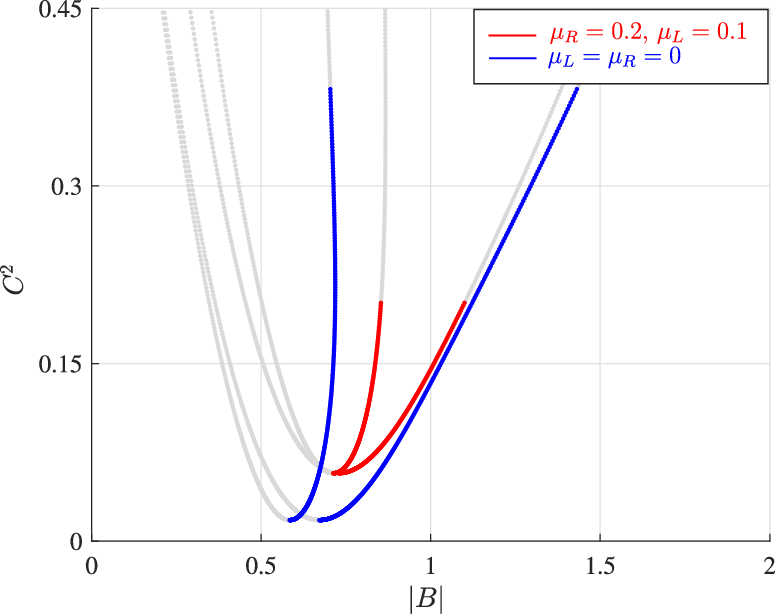}
		\caption{Stability region of the asymmetric $\mathcal{P}_2^1$-orbits for $r_L=0.56$, $r_R=0.76$ and $\tilde{g}=0.844$.}
		\label{Fig: friction_effect_stabilityrange}
	\end{minipage}\quad
	\begin{minipage}{0.475\textwidth}
		\includegraphics[width=\linewidth]{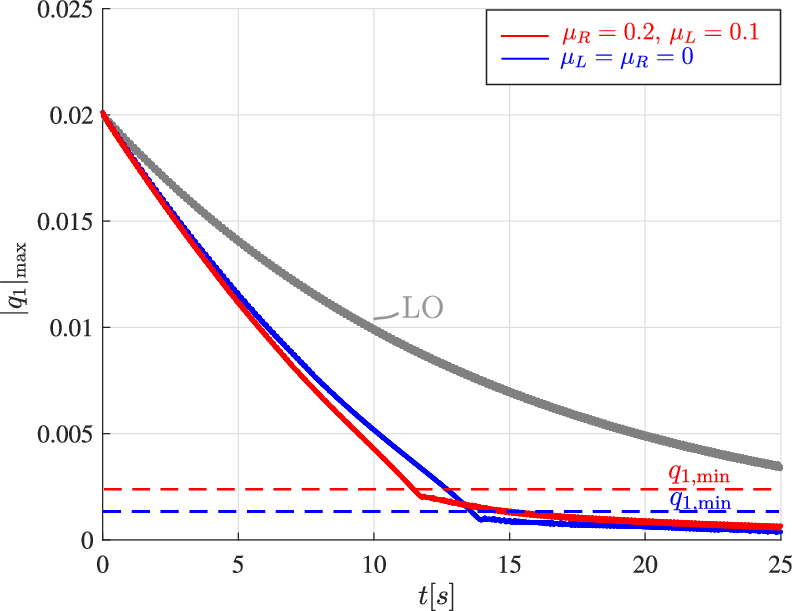}
		\caption{Amplitude of the free response of the main structure for $r_L=0.56$, $r_R=0.76$ and $\tilde{g}=0.844$.}
		\label{Fig: friction_effect_decayingamplitude}	
	\end{minipage}
\end{figure}

To further explore the effect of friction and asymmetry on stability and bifurcation behavior, a bifurcation diagram for an asymmetric case is shown in Figure (\ref{Fig: Bif_diagramm_asymm_final}). It reveals that asymmetric $\mathcal{P}_2^1$-orbits emerge at a fold bifurcation, with the absence of symmetric solutions. This result confirms that symmetric motion is only feasible with a symmetric configuration. Aside from this distinction, the overall qualitative behavior remains unchanged. The asymmetric branches lose stability at a period-doubling bifurcation, leading to $\mathcal{P}_4^2$-orbits, which remain stable until another period-doubling bifurcation occurs, beyond which $\mathcal{P}_8^4$-orbits are born. This period-doubling cascade continues, with a shrinking stability range after each bifurcation point. Similar to the symmetric case, higher friction coefficients raise the activation threshold of the VI-NES, as shown in Figure (\ref{Fig: friction_effect_stabilityrange}), where the threshold at which asymmetric $\mathcal{P}_2^1$-orbits appear is higher for the system with friction.
Apart from the absence of symmetric solutions, the overall qualitative behavior remains comparable to the frictionless case. At higher excitation amplitudes, additional periodic windows emerge, involving more impacts per cycle and exhibiting period-doubling cascades as well. As $C^2$ increases, the stability range for these motions expands, and the effect of friction becomes negligible, no longer affecting the width of the periodic windows.
This reduction in frictional effects can be seen in the formulation of the impact map, as friction-dependent terms become negligible for solutions with more impacts per cycle. Specifically, as $l$ increases, the time interval between consecutive impacts, defined in (\ref{eq: G LR })-(\ref{eq: G RL }) by the difference $\Psi_{n+i}-\Psi_{n+i-1}$ for $i=1 \cdots l$, decreases, thereby reducing the contribution of frictional terms in the computation of periodic solutions. Figure (\ref{Fig: friction_effect_decayingamplitude}) shows the envelope of the free response of the main structure from Figure (\ref{fig:model}), given by $|q_1|_{\max}\simeq b C$, for an asymmetric VINES with and without friction. At high amplitude levels, the decay rates of the response amplitude for both cases are nearly identical. However, as the amplitude decreases and the auxiliary mass undergoes fewer impacts per cycle, the effect of friction becomes significant, as expected. While its presence enhances the damping effect, it also causes the VI-NES activity to cease earlier due to its higher activation threshold.\\
The most striking observations revealed from the bifurcation diagrams across all three configurations —symmetric without friction, symmetric with friction, and asymmetric with friction— concern the stability characteristics of periodic orbits. In fact, symmetric periodic orbits (when present) lose stability due to symmetry-breaking bifurcations. In contrast, asymmetric periodic orbits become unstable at period-doubling bifurcations, with the stability ranges of the resulting branches shrinking compared to the initial ones. As the excitation amplitude $C$ increases, the frequency of period-doubling bifurcations within the same orbit cascade also increases, eventually ending the periodic windows and leading to chaos, as seen in the brute force diagrams. For symmetric periodic orbits of order $k=1$ (e.g.\ $\mathcal{P}_3^{1}$, $\mathcal{P}_4^{1}$,$\mathcal{P}_5^{1}$, $\cdots$), the corresponding stability ranges expand with higher excitation amplitudes $C$, as the number of impacts per period increases. These stability ranges are larger than those of asymmetric motions of the same order. In contrast, higher-order periodic orbits, such as $\mathcal{P}_3^8$ and $\mathcal{P}_5^{16}$, exhibit stability only within small ranges of excitation amplitudes.

%****************************************************************************
\section{Conclusion}
This work investigates the critical role of asymmetry and friction on the dynamics of the Vibro-Impact Nonlinear Energy Sink (VI-NES) using the impact map approach and path continuation. The findings confirm and extend previous results near the primary resonance, incorporating friction and generalizing the results to $1$:$k$ resonance. A detailed exploration of more complex periodic motions, involving different impact sequences, was conducted and their stability was analyzed using a linear perturbation technique. 
The results highlight several key observations: friction significantly affects the system at lower excitation levels, reducing the stability range of the two-impact motion.  This means that the analytically derived optimal activity range for the VI-NES, corresponding to the stable range of symmetric $\mathcal{P}_2^1$ orbits, is smaller in reality and only feasible for symmetric configurations. At higher excitation levels, however, the effect of friction diminishes, as bifurcations into solutions with more impacts per cycle occur. This behavior leads to the appearance of new periodic windows and a shift toward more complex motions. Notably, symmetric periodic orbits of order $k=1$ exhibit wider stability ranges at higher excitation amplitudes, while asymmetric periodic orbits show a reduction in the corresponding stability ranges following period-doubling bifurcations. This difference in behavior highlights the distinct impact of symmetry on the system's stability.
The analysis further reveals that friction increases the energy required to activate the VI-NES, while asymmetry significantly alters the bifurcation behavior of the solutions. As the excitation amplitude increases, the qualitative behavior of the asymmetric system subjected to friction approaches that of the frictionless case, with asymmetric motions exhibiting period-doubling cascades.
These findings are crucial for the use of a VI-NES in real-world applications as they provide a deeper understanding of the interplay between friction, asymmetry, and the dynamic behavior of the VI-NES. They also offer a solid foundation for future experimental validation and design optimization for the development of more efficient VI-systems.

\begin{acknowledgements}
This research was supported by the Deutsche Forschungsgemeinschaft (DFG).\\
Project number 402813361.
\end{acknowledgements}

\section*{Conflict of interest}
The authors declare that they have no conflict of interest.

% Non-BibTeX users please use


\begin{thebibliography}{}

	\bibitem{lu2018particle}
	Nucera, F., Vakakis, A. F., McFarland, D. M., Bergman, L. A., Kerschen, G.: Targeted energy transfers in vibro-impact oscillators for seismic mitigation. Nonlinear Dynamics, 50, 651-677 (2007).

	\bibitem{rahman2015performance}
	Rahman, M., Ong, Z. C., Chong, W. T., Julai, S., Khoo, S. Y.: Performance enhancement of wind turbine systems with vibration control: A review. Renewable and Sustainable Energy Reviews, 51, 43-54  (2015).
	
	\bibitem{chen2013tuned}
	Chen, J., Georgakis, C. T.: Tuned rolling-ball dampers for vibration control in wind turbines. Journal of Sound and Vibration, 332(21), 5271-5282  (2013).

	\bibitem{lu2018particle2}
	Lu, Z., Wang, Z., Masri, S. F., Lu, X.: Particle impact dampers: Past, present, and future. Structural Control and Health Monitoring, 25(1), e2058 (2018).
	


	\bibitem{gendelman2001transition}
	Gendelman, O. V.: Transition of energy to a nonlinear localized mode in a highly asymmetric system of two oscillators. Nonlinear dynamics, 25, 237-253 (2001).
	

	\bibitem{vakakis2001inducing}
	Vakakis, A. F.: Inducing passive nonlinear energy sinks in vibrating systems. J. Vib. Acoust., 123(3), 324-332 (2001).
	
		\bibitem{gendelman2001energy}
	Gendelman, O., Manevitch, L. I., Vakakis, A. F., M’Closkey, R.  Energy pumping in nonlinear mechanical oscillators: part I—dynamics of the underlying Hamiltonian systems. J. Appl. Mech., 68(1), 34-41 (2001).
	
	\bibitem{vakakis2001energy}
	Vakakis, A. F., Gendelman, O.: Energy pumping in nonlinear mechanical oscillators: part II—resonance capture. J. Appl. Mech., 68(1), 42-48 (2001).
	

	\bibitem{monjaraz2022prediction}
	Monjaraz-Tec, C., Kohlmann, L., Schwarz, S., Hartung, A., Gross, J., Krack, M. (2022). Prediction and validation of the strongly modulated forced response of two beams undergoing frictional impacts. Mechanical Systems and Signal Processing, 180, 109410.
	
	\bibitem{theurich2022predictive}
	Theurich, T., Vakakis, A. F., Krack, M.: Predictive design of impact absorbers for mitigating resonances of flexible structures using a semi-analytical approach. Journal of Sound and Vibration, 516, 116527 (2022).
	

	
	
	\bibitem{chatterjee1996impact}
	Chatterjee, S., Mallik, A. K., Ghosh, A.: Impact dampers for controlling self-excited oscillation. Journal of Sound and Vibration, 193(5), 1003-1014  (1996).
	
	
	\bibitem{wu2023targeted}
	Wu, Z., Paredes, M., Seguy, S.: Targeted energy transfer in a vibro-impact cubic NES: Description of regimes and optimal design. Journal of Sound and Vibration, 545, 117425  (2023).
	\bibitem{al2021comparison}
	AL-Shudeifat, M. A., Saeed, A. S.: Comparison of a modified vibro-impact nonlinear energy sink with other kinds of NESs. Meccanica, 56, 735-752 (2021).
	
	\bibitem{farid2021dynamics}
	Farid, M., Gendelman, O. V., Babitsky, V. I.:Dynamics of a hybrid vibro‐impact nonlinear energy sink. ZAMM‐Journal of Applied Mathematics and Mechanics/Zeitschrift für Angewandte Mathematik und Mechanik, 101(7), e201800341 (2021). 
	
	\bibitem{saeed2020rotary}
	Saeed, A. S., AL-Shudeifat, M. A., Vakakis, A. F., Cantwell, W. J.: Rotary-impact nonlinear energy sink for shock mitigation: analytical and numerical investigations. Archive of Applied Mechanics, 90, 495-521 (2020).
	
	\bibitem{al2013numerical}
Al-Shudeifat, M. A., Wierschem, N., Quinn, D. D., Vakakis, A. F., Bergman, L. A., Spencer Jr, B. F.: Numerical and experimental investigation of a highly effective single-sided vibro-impact non-linear energy sink for shock mitigation. International Journal of Non-linear Mechanics, 52, 96-109 (2013).
	%
	\bibitem{vakakis2008nonlinear}
	Vakakis, A. F., Gendelman, O. V., Bergman, L. A., McFarland, D. M., Kerschen, G., Lee, Y. S.: Nonlinear targeted energy transfer in mechanical and structural systems (Vol. 156). Springer Science and Business Media (2008).
	
	\bibitem{gendelman2015dynamics}
	Gendelman, O. V., Alloni, A.: Dynamics of forced system with vibro-impact energy sink. Journal of Sound and Vibration, 358, 301-314 (2015).
	
	\bibitem{gourc2015targeted}
	Gourc, E., Michon, G., Seguy, S., Berlioz, A.: Targeted energy transfer under harmonic forcing with a vibro-impact nonlinear energy sink: analytical and experimental developments. Journal of Vibration and Acoustics, 137(3), 031008 (2015).
	
	\bibitem{li2017dynamics}
	Li, T., Gourc, E., Seguy, S., Berlioz, A.: Dynamics of two vibro-impact nonlinear energy sinks in parallel under periodic and transient excitations. International Journal of Non-Linear Mechanics, 90, 100-110 (2017).
	
	\bibitem{qiu2019design}
	Qiu, D., Seguy, S., Paredes, M.: Design criteria for optimally tuned vibro-impact nonlinear energy sink. Journal of Sound and Vibration, 442, 497-513 (2019).
	
	\bibitem{li2017optimization}
	Li, T., Seguy, S., Berlioz, A.: Optimization mechanism of targeted energy transfer with vibro-impact energy sink under periodic and transient excitation. Nonlinear Dynamics, 87, 2415-2433 (2017).
	
	\bibitem{li2025effectiveness}
	Li, H., Li, S., Zhang, Z., Xiong, H., Ding, Q.. Effectiveness of vibro-impact nonlinear energy sinks for vibration suppression of beams under traveling loads. Mechanical Systems and Signal Processing, 223, 111861 (2025).
	
	
	\bibitem{theurich2019effects}
	Theurich, T., Gross, J., Krack, M.: Effects of modal energy scattering and friction on the resonance mitigation with an impact absorber. Journal of Sound and Vibration, 442, 71-89 (2019).
	
	\bibitem{pennisi2017experimental}
	Pennisi, G., Stephan, C., Gourc, E., Michon, G.: Experimental investigation and analytical description of a vibro-impact NES coupled to a single-degree-of-freedom linear oscillator harmonically forced. Nonlinear Dynamics, 88, 1769-1784 (2017).
	
	\bibitem{li2021importance}
	Li, H., Li, A., Zhang, Y.: Importance of gravity and friction on the targeted energy transfer of vibro-impact nonlinear energy sink. International Journal of Impact Engineering, 157, 104001 (2021).
	\bibitem{wang2016numerical}
	Wang, J., Wierschem, N., Spencer Jr, B. F., Lu, X.: Numerical and experimental study of the performance of a single‐sided vibro‐impact track nonlinear energy sink. Earthquake Engineering  Structural Dynamics, 45(4), 635-652 (2016).
	
	\bibitem{leine2012global}
	Leine, R. I., Heimsch, T. F.: Global uniform symptotic attractive stability of the non-autonomous bouncing ball system. Physica D: Nonlinear Phenomena, 241(22), 2029-2041 (2012).
	
	\bibitem{youssef2021complete}
	Youssef, B., Leine, R. I.: A complete set of design rules for a vibro-impact NES based on a multiple scales approximation of a nonlinear mode. Journal of Sound and Vibration, 501, 116043 (2021).
	
	
	\bibitem{bapat1985single}
	Bapat, C. N., Sankar, S.: Single unit impact damper in free and forced vibration. Journal of Sound and Vibration, 99(1), 85-94 (1985).

\bibitem{liu2023maps}
	Liu, R., Kuske, R., Yurchenko, D.  Maps unlock the full dynamics of targeted energy transfer via a vibro-impact nonlinear energy sink. Mechanical Systems and Signal Processing, 191, 110158 (2023).
	\bibitem{leine2007stability}
Leine, R. I.,  van de Wouw, N.: Stability and Convergence of Mechanical Systems with Unilateral Constraints (Lecture Notes in Applied ans Computational Mechanics, Vol. 36). Springer Science  Business Media (2007).

\end{thebibliography}
\end{document}